\documentclass[letter,times,twocolumn,tighten]{aastex62}
\received{January 1, 2018}
\revised{January 7, 2018}
\accepted{\today}
\submitjournal{ApJL}

\shorttitle{A Doppler-flip in HD\,100546} 
\shortauthors{Casassus \& Perez}

\usepackage{natbib}  
\usepackage{epsfig,graphicx,graphics}
\usepackage{amsmath}	
\usepackage{amssymb}	

\begin{document}


\title{Kinematic detections of protoplanets: a Doppler-flip in the
  disk of HD\,100546.}

\correspondingauthor{Simon Casassus}
\email{simon@das.uchile.cl}

\author[0000-0002-0433-9840]{Simon Casassus}
\affiliation{Departamento de Astronom\'{\i}a, Universidad de Chile, Casilla 36-D, Santiago, Chile}

\author[0000-0003-2953-755X]{Sebasti\'an P\'erez}
\affiliation{Universidad de Santiago de Chile, Av. Libertador Bernardo
  O'Higgins 3363, Estaci\'on Central, Santiago, Chile}




\begin{abstract}

 
Protoplanets and circumplanetary disks are rather elusive in their
thermal IR emission. Yet they are cornerstone to the most popular
interpretations for the protoplanetary disks structures observed in
the gas and dust density fields, even though alternative theories
exist. The gaseous velocity field should also bear the imprint of
planet-disk interactions, with non-Keplerian fine structure in the
molecular line channel maps. Such kinks or wiggles are affected by the
optical depth structure and synthesis imaging limitations, but their
detail could in principle be connected to the perturber by comparison
with hydrodynamical simulations. These predictions appear to have been
observed in HD\,163296 and HD\,97048, where the most conspicuous
wiggles are interpreted in terms of embedded planets. The velocity
centroid maps may allow for more robust indirect detections of
embedded planets.  The non-Keplerian velocity along the planetary
wakes undergoes an abrupt sign reversal across the protoplanet.  After
subtraction of the disk rotation curve, the location of the perturber
should be identifiable as a Doppler-flip in velocity centroid
maps. Here we improve our rotation curves in an extension to disks
with intermediate inclinations, which we apply to deep and fine
angular resolution CO isotopologue datasets. Trials in HD\,163296 and
in HD\,97048 yield non-detections. However, in HD\,100546 we pick-up a
conspicuous Doppler-flip, an important part of which is likely due to
radial flows. Its coincidence with a fine ridge crossing an annular
groove inside the continuum ring suggests a complex dynamical
scenario, in which the putative protoplanet might have recently
undergone pebble accretion.


\end{abstract}



\section{Introduction} \label{sec:intro}

Direct detections of protoplanets and their circum-planetary accretion
disks (CPDs) are difficult in their thermal IR emission. Only one
example appears to hold scrutiny
\citep[i.e. PDS70b,][]{Keppler2018A&A...617A..44K}. Yet most models for
the structures observed in protoplanetary disks involve planet-disk
interactions: giant planets evacuate radial gaps, launch spiral arms,
and trigger crescent-shaped pile-ups of mm-grains. The gas and dust
density fields are thus appealing proxies of the location and mass of
embedded bodies \citep[e.g.][]{DongFung2017ApJ...835..146D}. The
planetary origin of such density structures is debated, however, as
exemplified by the alternative scenarios in HL\,Tau and TW\,Hya
\citep[e.g. ][and references therein]{Dong2018ApJ...866..110D}.

Molecular-line kinematics provide an alternative approach to the
identification of protoplanets, as the gaseous velocity field also
bears the imprint of planet-disk interactions.  We have linked
embedded planets with wiggle-shaped deviations from sub-Keplerian
rotation in channel maps
\citep[][]{Perez2015ApJ...811L...5P}. However, such wiggles are
ubiquitous in channel maps and can also be due to noise or systematics
from synthesis imaging, or from structure in the underlying gas and
dust optical depth. Hydrodynamical simulations are required to infer
the location of embedded planets from the observed channel map
wiggles. These predictions appear to match the data in HD\,163296 and
in HD\,97048, where the most conspicuous wiggles are interpreted in
terms of the location and mass of the perturber
\citep[][]{Pinte2018ApJ...860L..13P, Pinte2019arXiv190702538P}. In a
companion Letter (Perez et al., 2019, ApJL, submitted - hereafter
paper\,I) we report on another conspicuous wiggle in HD\,100546.

The location of the perturbing body should be identifiable in velocity
centroid maps $ v_\circ(\vec{x}) = \langle vI\rangle = \int dv v I_v/
\int I_v dv$, where $I_v(\vec{x})$ are the molecular line channel maps
as a function of line-of-sight velocity $v$.  In hydrodynamical
simulations, and typically for the disk midplane, the radial and
azimuthal velocity deviations after subtraction of the sub-Keplerian
background flow, are highest along the planetary wakes, and abruptly
change sign at the location of the planet \citep[e.g. see Fig.\,1
  in][]{Perez2018MNRAS.480L..12P}.  At non-zero inclination this
velocity reversal should be observable as a Doppler-flip, provided
that the azimuthally symmetric background is adequately
subtracted. Away from the midplane, the planetary wakes have
significant vertical and radial motion which increases with altitude
\citep{Zhu2015}. Most of the material interacting with a giant planet
falls onto the midplane from high latitudes via meridional flows
\citep{Morbidelli2014, Dong2019}, including both radial and vertical
velocity components. We thus designed a technique to filter the
velocity centroid image and pin-point the location of the perturber,
yielding favorable predictions for 5\, to 10\,M$_\mathrm{jup}$ bodies
given current ALMA capabilities \citep[][]{Perez2018MNRAS.480L..12P}.

The rotation curve of a disk, $\tilde{v}_\phi(r) = \int dz d\phi \,
\rho v_\phi / \int dz d\phi \, \rho$, in cylindrical coordinates,
represents the mass-averaged tangential velocity in the plane of the
disk as a function of radius.  Among other uses in astrophysics,
$\tilde{v}_\phi(r)$ allows a measure of the stellar mass in
circumstellar disks. Here we are interested in building the rotation
curve of protoplanetary disks with the goal of detecting deviations
from azimuthal symmetry. We improve our method by taking azimuthal
averages on conical surfaces (Section\,\ref{sec:conictransform}), and
report applications to observations (Sec.\,\ref{sec:applications})
with emphasis on HD\,100546, where we pick-up a conspicuous
Doppler-flip.



\section{Rotation curves and orientation of flared
 disks from  velocity centroid maps} \label{sec:conictransform}



Dynamical stellar masses and measurements of $\tilde{v}_\phi(r)$ are
usually obtained with forward-modeling of molecular-line
data. Parametric radiative transfer (RT) modeling, with prescribed
$\tilde{v}_\phi(r)$, can be compared against observations of
$I_v(\vec{x})$ \citep[e.g.][]{Czekala2015ApJ...806..154C}. Such
dynamical masses are indeed within 5--10\% of independent estimates
for close binaries
\citep[][]{Rosenfeld2012ApJ...759..119R,Czekala2015ApJ...806..154C,
  Czekala2016ApJ...818..156C}. However, the inferred
$\tilde{v}_\phi(r)$ may not necessarily follow the details of the
rotation curve, for instance because of the vertical velocity
structure. Rotation curves have recently been measured empirically by
noting that the tangential velocity component and height over the disk
midplane can be solved for at two specific locations in the sky (for
each of the near and far sides of the disk), given knowledge of disk
orientation \citep[][]{Pinte2018A&A...609A..47P}. This local datum for
the disk height (or aspect radio $h$) has been extrapolated to
deproject a circle at constant height, and so fit for
$\tilde{v}_\phi(r)$ \citep[][]{Teague2018ApJ...860L..12T}. However,
uncertainties in the local measurement of $h$, if it is itself noisy
or perturbed by local deviations from axial symmetry, may propagate in
the azimuthal averages when extrapolated to all azimuths.




Here the measurement of $\tilde{v}_\phi(r)$ is not driven by the
dynamical mass, but rather by the deviations in $v_\circ(\vec{x})$
from the azimuthally-averaged background
$\tilde{v}_\circ(\vec{x})=\tilde{v}_\phi(r)\cos(\phi)$, where the
origin of $\phi$ coincides with the disk position angle (PA). We infer
the bulk flow $\tilde{v}_\circ(\vec{x})$ directly from the observed
velocity centroid map $v_\circ(\vec{x})$ and its uncertainty
$\sigma^2_\circ(\vec{x})$, extracted from $I_v(\vec{x})$ using
single-Gaussian fits to each line-of-sight spectrum (including a
linear baseline for the continuum). Our main assumption is that the
surface in the disk where the molecular line emission originates,
i.e. the unit-opacity surface for optically thick lines, can be
represented by a double cone.  The free-parameters are reduced
compared to forward-modeling, to only ($i$, PA, $h$). Disk flaring or
radial variations in $h$ can be measured by binning the optimization
in radius.


We consider three coordinates systems: $\mathcal{S}$ represents the
sky frame, orientated with $(x,y,z)$ Cartesian coordinates, and where
$y$ corresponds to North. $\mathcal{S}^\prime$ is also parallel to the
sky, but its $y^\prime$ axis coincides with the disk
PA. $\mathcal{S}^{\prime\prime}$ is the rotation of
$\mathcal{S}^\prime$ about axis $\hat{y}^\prime$ by the inclination
angle $i$, so that the $z^{\prime\prime}=0$ coincides with the disk
midplane.

If all of the emission originates from the near side, we have a
bijection between the line of sight $\vec{x}$, and the polar
coordinate of its intersection with the surface of the cone
representing the disk surface (defined as the surface of unit
opacity). Given a point on the cone, with an opening angle $\psi=\arctan(h/r)$
above the disk midplane, and with cylindrical coordinates $r,\phi$ in
$\mathcal{S}^{\prime\prime}$, we can compute the sky coordinates in
$\mathcal{S}^\prime$ with the following conical transform
$\vec{x}^\prime = \vec{f}_{\mathrm{PA},i,+\psi}(r,\phi)$ \citep[see
  also][]{Rosenfeld2013ApJ...774...16R, Isella2018ApJ...869L..49I}:
\begin{eqnarray}
    x^\prime& = &r  \sin(\phi) / \cos(i) + (h - r \sin(\phi) \tan(i))\sin(i),  \label{eq:xforward}\\ 
    y^\prime& = &r \cos(\phi). \label{eq:yforward}
\end{eqnarray}
Here we use a linear law for $h(r) = \tan(\psi) r$.  We can invert Eqs.~\ref{eq:xforward} and \ref{eq:yforward} to obtain
$(r,\phi) = \vec{f}^{-1}_{\mathrm{PA},i,+\psi}(x^\prime,y^\prime)$, by
noting that $r$ is the root of:
\begin{equation}
  \frac{y^{\prime\, 2}}{r^2} + \frac{(x^{\prime}-h(r)\,\sin(i))^2}{r^2\cos^2(i)} = 1.
\end{equation}
Having solved for $r$ we obtain $\phi$ with
\begin{equation}
 \cos(\phi) = y^\prime / r.
\end{equation}


%




Given (PA, $i$, $\psi$) we resample the observed centroid
$v_\circ(\vec{x})$ with the near-side conical transform
$\vec{f}_{\mathrm{PA},i,+\psi}$:
\begin{equation}
  v_{\circ+}(r,\phi) \equiv v_\circ\left(\vec{f}^{-1}_{\mathrm{PA},i,+\psi}(r,\phi )\right), 
\end{equation}
where the $+$ sign in $v_{\circ+}$ indicates the near-side conical
surface (note that for $i > 90\,$deg, the near side corresponds
to $\psi < 0$). We then take azimuthal averages,
\begin{equation}
  v^m_{\circ+}(r_l,\phi_k) = \cos(\phi_k) \tilde{v}_\phi(r_l) + v_s,  \label{eq:vmodel+}
\end{equation}
in discretized  polar coordinates, by fitting for the rotation curve
$\tilde{v}_\phi(r_l)$ in a least-squares sense with 
\begin{equation}
\chi^2_{\tilde{v}} = \sum_k  w(r_l,\phi_k) ( v_{\circ+}(r_l,\phi_k)- \tilde{v}_\phi(r_l)  \cos(\phi_k) - v_s)^2, 
\end{equation}
with weights $w(r_l,\phi_k)=1/\sigma^2_\circ(r_l,\phi_k)$. When the
systemic velocity $v_s$ is known,
\begin{equation}
  \tilde{v}_\phi(r_l) = \frac{\sum_k w(r_l,\phi_k) [v_{\circ+}(r_l,\phi_k) -
    v_s ] \cos(\phi_k)  }{\sum_k w(r_l,\phi_k) \cos^2(\phi_k)}.
\end{equation}
If $v_s$ is not known we can fit for $v_s(r_l)$ and
$\tilde{v}_\phi(r_l)$ simultaneously, fix $v_s$ to the median value
and its uncertainty to its standard deviation, and then recompute
$\tilde{v}_\phi(r_l)$. We resample $v^m_{\circ+}(r,\phi)$ with
$\vec{x}^\prime = \vec{f}_{\mathrm{PA},i,+\psi}(r,\phi)$ and rotate
back to $\mathcal{S}$ to obtain the sky map for the
azimuthally-averaged centroid, $v^m_{\circ+}(\vec{x})$.



We optimize $i$, PA and $\psi$ so as to maximize the log-likelihood
function, $-0.5\chi^2$, with
\begin{equation}
 \chi^2 =  \frac{1}{N_\mathrm{corr}}\sum_{l=l1}^{l2} \sum_{k=0}^{N_\phi} w(r_l,\phi_k) (v_\circ(r_l,\phi_k) - v^m_{\circ+}(r_l,\phi_k))^2.
\end{equation}
The double sum extends over pixels in polar coordinates, within an
interval in radius between $r_1$ and $r_2$, and over all azimuths.
$N_\mathrm{corr}$ is the number of pixels in a beam, and approximately
corrects $\chi^2$ for correlated datapoints.

Disk orientation may change with radius; even in the absence of
warping the disk aspect ratio will vary. Passive disks are expected to
be flared, but the height of the $^{12}$CO unit opacity surface will
decrease with radius, so the trend in $\psi(r)$ is difficult to
anticipate. We divide the full radial extension of the disk in $M$
(overlapping) radial bins $\{[r_{1j}, r_{2j}]\}_{j=1}^{M}$, thus
defining radial regions $\{ \Theta_j(\vec{x}^{\prime}) \}_{j=1}^{M}$,
where $\Theta_j(\vec{x}^{\prime})=1$ if $r=f_r^{-1}(\vec{x}^{\prime})
\in [r_{1j}, r_{2j}]$, and $\Theta_j(\vec{x}^{\prime})=0$
otherwise. We take azimuthal averages of $v_\circ(\vec{x})$ in each
region to produce $\{ v^m_{j}(\vec{x})\}_{j=1}^M$.  The azimuthally
averaged centroid map combined over all regions is
\begin{equation}
  v^m_{\circ R}(\vec{x}) = \frac{\sum_{j=1}^M
    v^m_{\circ+}(\vec{x}) \Theta_j(\vec{x})} {\sum_{j=1}^M
    \Theta_j(\vec{x})}.
\end{equation}
In the applications below, the optimal orientation for the whole
radial extension is $i_\circ$, PA$_\circ$ and $\psi_\circ$, while the
orientation profile is $i(r)$, PA$(r)$ and $\psi(r)$.


At finite inclinations parts of the far-side will contribute to
$v_\circ(\vec{x})$, with a weight $\mu(\vec{x})$ that depends on the
technique used to measure the velocity centroid. Including the
far-side yielded small improvements in $\chi^2$, but complicates the
interpretation of the velocity deviations, so we opted to present
results with $\mu(\vec{x})\equiv 1$ only.

\section{Applications} \label{sec:applications}

\subsection{HD\,100546}

\begin{figure*}
\centering
\includegraphics[width=0.8\textwidth]{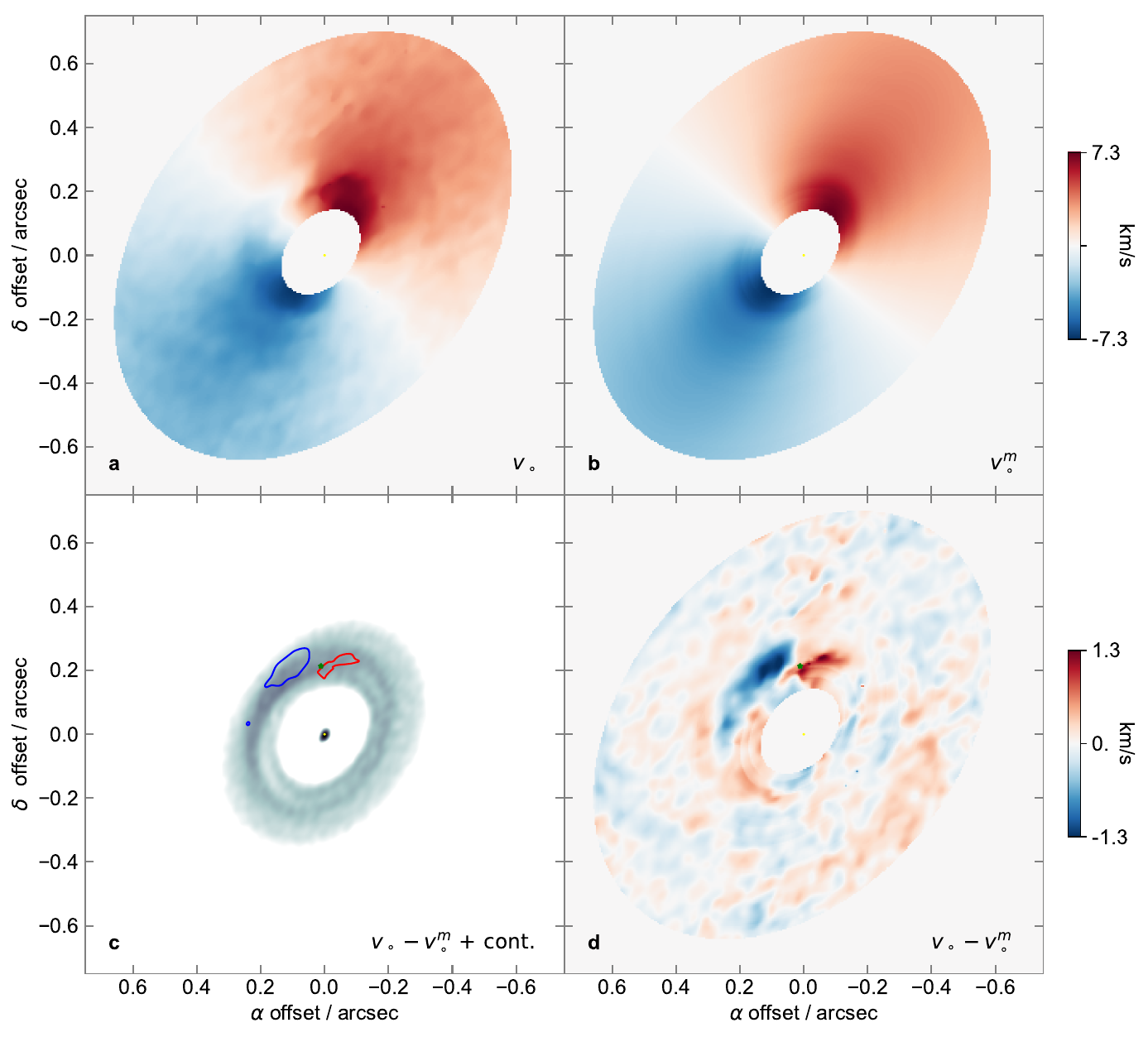}
\includegraphics[width=0.8\textwidth]{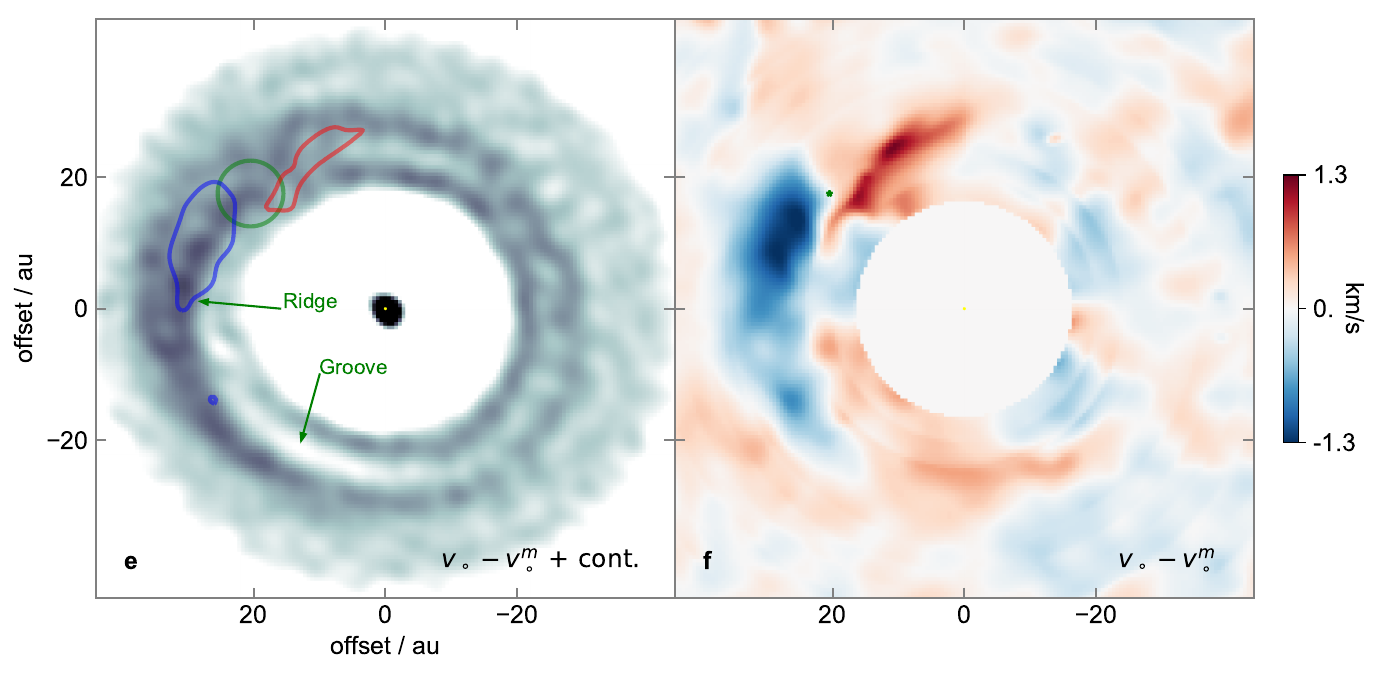}
\caption{Doppler-flip in HD\,100546. {\bf a}: $v_\circ$, observed
  $^{12}$CO(2-1) line centroid (paper I, $0\farcs076 \times
  0\farcs057$ beam). {\bf b:} $v_\circ^m$, azimuthal average of the
  line centroid.  {\bf c:} Contours for $v_\circ - v_\circ^m$, taken
  at 60\% peak in both red and blue, overlaid on the 225\,GHz
  continuum (paper\,I). The green asterisk indicates the sky position
  of the putative protoplanet, assuming that it is embedded in the
  disk midplane.  {\bf d:} $v_\circ - v_\circ^m$, non-axially
  symmetric velocity field. {\bf e:} face-on view of $v_\circ -
  v_\circ^m$ in contours, overlaid on an unsharp mask of the a
  deconvolved continuum image (paper I, angular resolution
  $\sim$$20\times13$~mas$^2$ beam); $x-$ and $y-$ axis show offset in au,
  for a distance of $110.02$\,pc. The green circle is centered on the
  position of the putative protoplanet.  {\bf f:} face-on view of
  $v_\circ - v_\circ^m$.
  \label{fig:mom1_HD100546} }
\end{figure*}

In paper\,I we report on a conspicuous wiggle in $^{12}$CO(2-1) ALMA
observations\footnote{Cycle\,4 project 2016.1.00344.S; we used CASA
  task {\tt tclean} and Briggs weighting, with a robustness parameter
  of 1.0, to produce 0.5\,km\,s$^{-1}$ channel maps with a beam
  $0\farcs076 \times 0\farcs057$ elongated towards $-34\deg$} of
HD\,100546. We apply our velocity filter to the velocity centroid map
extracted from these data.  Fig.\,\ref{fig:mom1_HD100546} summarizes
the result of an optimization of disk orientation and rotation
curve. The difference between observations and azimuthally-averaged
model, $v_\circ - v_\circ^m$, in Fig.\,\ref{fig:mom1_HD100546}\,d
shows a Doppler-flip related to that wiggle, with a total amplitude of
2.6\,km\,s$^{-1}$ where the rotation curve is $\tilde{v}_\phi
\sim$6\,km\,s$^{-1}$. We can also identify spiral arms in $v_\circ -
v_\circ^m$, some of which have previously been detected in the IR
\citep[][]{Boccaletti2013A&A...560A..20B,Garufi2016A&A...588A...8G}.

The morphology of the flip in HD\,100546
(Fig.\,\ref{fig:mom1_HD100546}) is similar to that predicted for
disk-planet interactions \citep[][]{Perez2018MNRAS.480L..12P},
especially in its azimuthal extension and in the sign of the velocity
deviations. Along a radius of $\sim0\farcs25$ we find several sign
oscillations in $v_\circ - v_\circ^m$, but the blue and red extrema
stand out. The main blue-shifted arc extends across the disk minor
axis, which suggests that an important fraction of the non-Keplerian
velocity component in the flip is either vertical or radial. The red
arc appears to terminate along the disk major axis, along which
$v_\circ - v_\circ^m$ is in general small, which could either be a
coincidence or reflect that the velocity deviations are more radial
than vertical.  HD\,100546 is rotating anti-clockwise, so that the
blue-shifted arc is probably infalling.  However, well outside or
inside of the blue arc no particular trend in $v_\circ - v_\circ^m$ is
evident, apart perhaps for somewhat smaller values along the minor
axis, suggesting that azimuthal deviations may still dominate in the
rest of the disk.

If the flip is due to a compact body, the perturber is probably found
in the disk midplane. Figs.\,\ref{fig:mom1_HD100546}d and e provide
face-on views of the velocity field corresponding to the radial
orientation profile. The location of the putative protoplanet, at the
center of the flip, is roughly $(20.5\pm5,17.5\pm5)$\,au, with errors
that indicate the radial extent of the flip. For approximate
consistency with the CO orientation profile, we assumed a flat disk in
order to deproject the continuum image with the global orientation of
the CO disk ($i_\circ$, PA$_\circ$, $\psi_\circ$) measured from
$0\farcs15$ to $0\farcs75$ (see below).  This does not necessarily
minimize the eccentricity of the continuum ring. The position on the
sky of the putative protoplanet corresponds to $(0\farcs 01\pm 0\farcs
04,0\farcs 21\pm 0\farcs 04)$, and falls under the red part of the
flip because of the projection.


The putative protoplanet appears to be embedded within the continuum
ring\footnote{the 225\,GHz continuum image used in
  Fig.\,\ref{fig:mom1_HD100546} is a deconvolution obtained with the
  {\sc uvmem} package \citep[][]{Carcamo2018A&C....22...16C}, with an
  approximate angular resolution of $20\times13$~mas (see
  paper\,I)}.  This is surprising because a single giant planet on a
fixed orbit, or undergoing type\,{\sc ii} migration, is known to
evacuate a gap \citep[][]{LinPapaloizou1980MNRAS.191...37L,
  GoldreichTremaine1980ApJ...241..425G}.  Also, the pitch-angle of the
planetary wakes has the opposite sign from the non-migrating
predictions \citep[][]{Perez2018MNRAS.480L..12P}. This puzzling
configuration may perhaps be due to a migration scenario in which the
giant planet causing the flip would have entered a radial pressure
maximum at the edge of the deep cavity caused by closer-in and massive
planets, such as the companion proposed by \citet[][thought to orbit
  roughly at the edge of the gaseous
  cavity]{Brittain2014ApJ...791..136B}. Or perhaps this giant planet
has only recently undergone runaway pebble accretion
\citep[e.g.][]{Liu2019A&A...624A.114L}, while still embedded in the
dense dust trap that led to core accretion.  The morphology of the
continuum also points at a complex dynamical scenario, with
overlapping ellipsoidal rings having different periapses, or
converging spiral arms (a constant disk inclination is favored by the
kinematics, see Fig.\,\ref{fig:rotorient_HD100546} below). Perhaps the
body responsible for the Doppler-flip is also causing the groove
indicated in Fig.\,\ref{fig:mom1_HD100546}e, and transient aerodynamic
interactions may perhaps collect the mm-emitting dust along the narrow
continuum ridge.



It may be tempting to associate the flip with an anticyclonic vortex
but the center of the flip is not associated with a continuum peak
\citep[as in][their Fig.7 case C]{HuangP2018ApJ...867....3H}. Also, if
the flip is mainly associated to radial flows, then it would be
rotating counter-clockwise, which would correspond to a cyclonic
vortex that would not survive Keplerian shear. By contrast the
Doppler-flip seems to embrace the continuum ridge upstream of the
continuum peak. If the CO velocity field reflects that of the
underlying continuum, the continuum ridge appears to flow towards the
center of the flip. Instead, if the local dust concentration causing
the continuum peak at PA$\sim$45\,deg is rotating slower than the gas
\citep[as would a vortex, e.g.][]{ZB2016MNRAS.458.3918Z}, then the gas
is overtaking the dust clump at the location of the Doppler-flip. In
this case the gas would be flowing over the ridge as if it were an
airfoil, with a vertical velocity component towards the observer.


The optical depth structure of the disk can produce wiggles in channel
maps, even with a purely azimuthal velocity field. However, such
wiggles cancel along the disk minor axis, where the associated line of
sight velocity is of course constant at the systemic value. The
overlap of the blue arc in the Doppler flip with the disk minor axis
cannot be explained by such optical depth effects. Still, part of the
velocity deviations away from the minor axis could correspond to
optical depth effects in the ring, with an increased line optical
depth and a strong and lopsided continuum.  We have tested for this
possibility with RT modeling computed using the RADMC3D package
\citep[][]{RADMC3D0.41}, including an exaggeratedly thick and compact
clump. The largest deviations $v_\circ - v_\circ^m$ for the emergent
$^{12}$CO(2-1) amount to 0.1\,km\,s$^{-1}$, such biases are thus small
compared to the magnitude of the observed flip. This RT model also
confirms that the disk orientation is recovered, within $\sim$0.2\,deg
for inclination.

%
%



We optimized the disk orientation in $M=15$ radial bins from
$0\farcs15$ to $0\farcs75$ to obtain the radial profiles $i(r)$,
PA$(r)$ and $\psi(r)$ shown in Fig.\,\ref{fig:rotorient_HD100546}. The
PA and inclination $i$ both show small variations (relative to their
typical range of values), without a systematic trend as would be the
case for a warp. By contrast $\psi$ varies considerably and
systematically with radius.  Motivated by the apparent lack of a
detectable warp, the azimuthal averages and face-on deprojections in
Fig.\,\ref{fig:mom1_HD100546} assume that the disk is not warped and
has constant $i(r)=i_\circ$ and PA$(r)$=PA$_\circ$. The variations in
$i(r)$ and PA$(r)$ yields essentially the same result, with a somewhat
smaller Doppler-flip amplitude (of 2.5\,km\,s$^{-1}$), and small but
unaesthetic discontinuities associated to the radial bins. The best
fit global orientation from $r_1=0\farcs15$ to $r_2=0\farcs75$ is
$i_\circ = 45.77\pm 1.55\,$deg, PA$_\circ=321.2 \pm1.1\,$deg,
$\psi_\circ=9.3\pm2.5\,$deg, where the uncertainties correspond to the
standard deviations of each of the radial profiles.  The Gaussian fits
used to extract the velocity profiles also yield an error map
$\sigma(v_\circ(\vec{x}))$. We estimated the resulting uncertainties
on the orientation profile using the {\tt emcee} package
\citep[][]{emcee2013PASP..125..306F}, but these `thermal' errors are
rather small. Instead, the uncertainty on the orientation profile are
mostly systematic and connected to the simplifications in our
averaging procedure, which is why we take statistics on the radial
profiles.  The systemic velocity is $5.68\pm 0.03\,$km\,s$^{-1}$.






\begin{figure}
\centering
\includegraphics[width=\columnwidth]{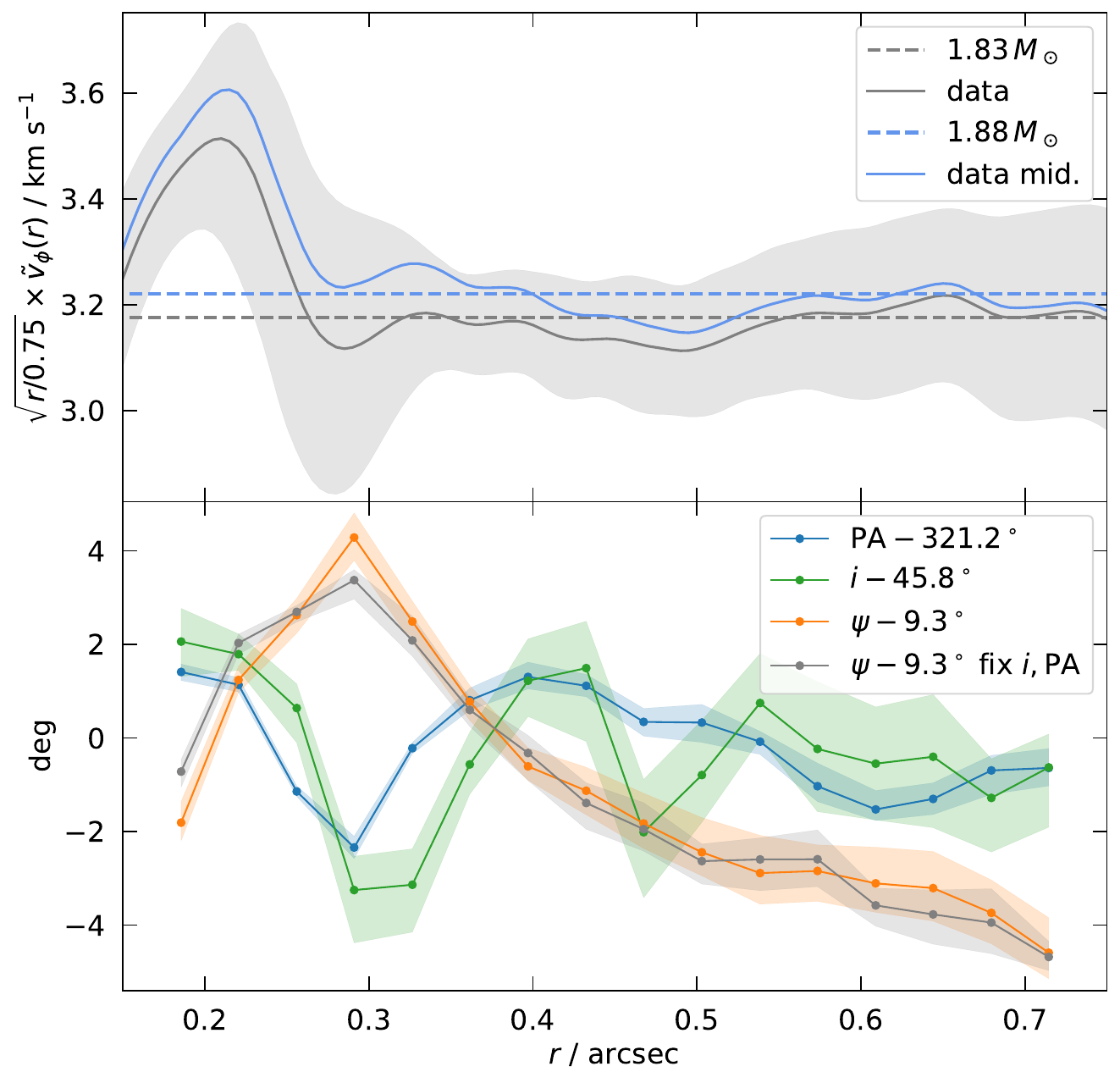}
\caption{Top: Rotation curve of HD\,100546 for the global orientation
  $i_\circ$, PA$_\circ$, $\psi_\circ$. The thin grey line traces
  $\tilde{v}_\phi(r)$, and the vertical extension of the shaded area
  corresponds to twice the scatter of $v_\circ(\phi) -
  v^m_\circ(\phi)$ (this scatter is $\sim10\times$ the error on
  $\tilde{v}_\phi(r)$). Comparison Keplerian profiles are shown in
  dashed lines.  Bottom: Orientation of HD\,100546, with radial
  variations in the deviations of PA, $i$, and $\psi$ from their value
  for the whole radial domain. We also plot $\psi(r)$ when fixing
  $i=i_\circ$ and PA=PA$_\circ$.  \label{fig:rotorient_HD100546} }
\end{figure}


The rotation curve $\tilde{v}_\phi(r)$ for the global orientation
$i_\circ$, PA$_\circ$, $\psi(r)$, shown in
Fig.\,\ref{fig:rotorient_HD100546}, is readily comparable with
Keplerian profiles \citep[ignoring corrections due to hydrostatic
  support, e.g.][]{Rosenfeld2013ApJ...774...16R,
  Teague2018ApJ...860L..12T}, with $v_K(r) = \sqrt{GM_\star / (r\times
  d)}$ for a distance $d$. We checked that the impact of calculating
$\tilde{v}_\phi(r)$ with the radial profiles $i(r)$, PA$(r)$,
$\psi(r)$ is small (below 1\%). If we assume a constant
$\tilde{v}_\phi(r)$ independent of height over the midplane, the best
fit mass from $0\farcs15$ to $0\farcs75$ is $M_\star =
1.83\pm0.01\,M_\odot$, as measured by minimizing
\begin{equation}
\chi^2 = \sum_r \frac{(v_K(r) - \tilde{v}_\phi(r))^2}
{\sigma^2(\tilde{v}_\phi)},
\end{equation}
and for a distance of 110.02\,pc \citep[][]{Gaia2018A&A...616A...1G}.
The uncertainties in $M_\star$ are estimated using the error on the
mean $\sigma(\tilde{v}_\phi)$, which is smaller than the scatter
plotted in Fig.\,\ref{fig:rotorient_HD100546} by a factor $\sim
1/\sqrt{N_b} \sim 1/10$, where $N_b$ is the number of independent data
points (i.e the number of clean beams along a fixed radius). The
standard deviation of $M_\star$ when fitting in narrow radial bins is
$\sigma(M_\star) = 0.12\,M_\odot$, we adopt this uncertainty to
account for systematics. Vertical Keplerian shear, in the extreme case
of no vertical viscosity or magnetic coupling, and ignoring radial
hydrostatic support, yields midplane azimuthal velocities
$\tilde{v}_\phi(r) (1+h^2)^{3/4}$. This is the case labeled `data
mid.' in Fig.\,\ref{fig:rotorient_HD100546}, with a slightly higher
$M_\star =1.88 \pm 0.13\,M_\odot$.  The deviations of
$\tilde{v}_\phi(r)$ from $v_K(r)$ are much larger than in this
midplane extrapolation, and are probably due to the radial pressure
gradient. The non-Keplerian rotation curve is particularly pronounced
inside a radius of $0\farcs3$, which could reflect a pressure bump
under the continuum ring at the edge of the cavity.




\subsection{HD\,97048 and HD\,163296}

\begin{figure}
\centering
\includegraphics[width=\columnwidth]{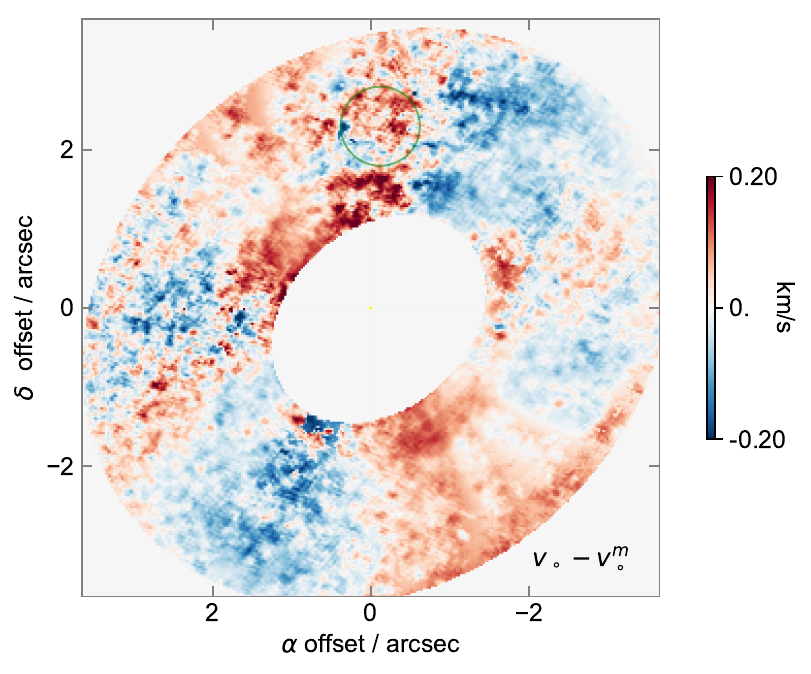}
\includegraphics[width=\columnwidth]{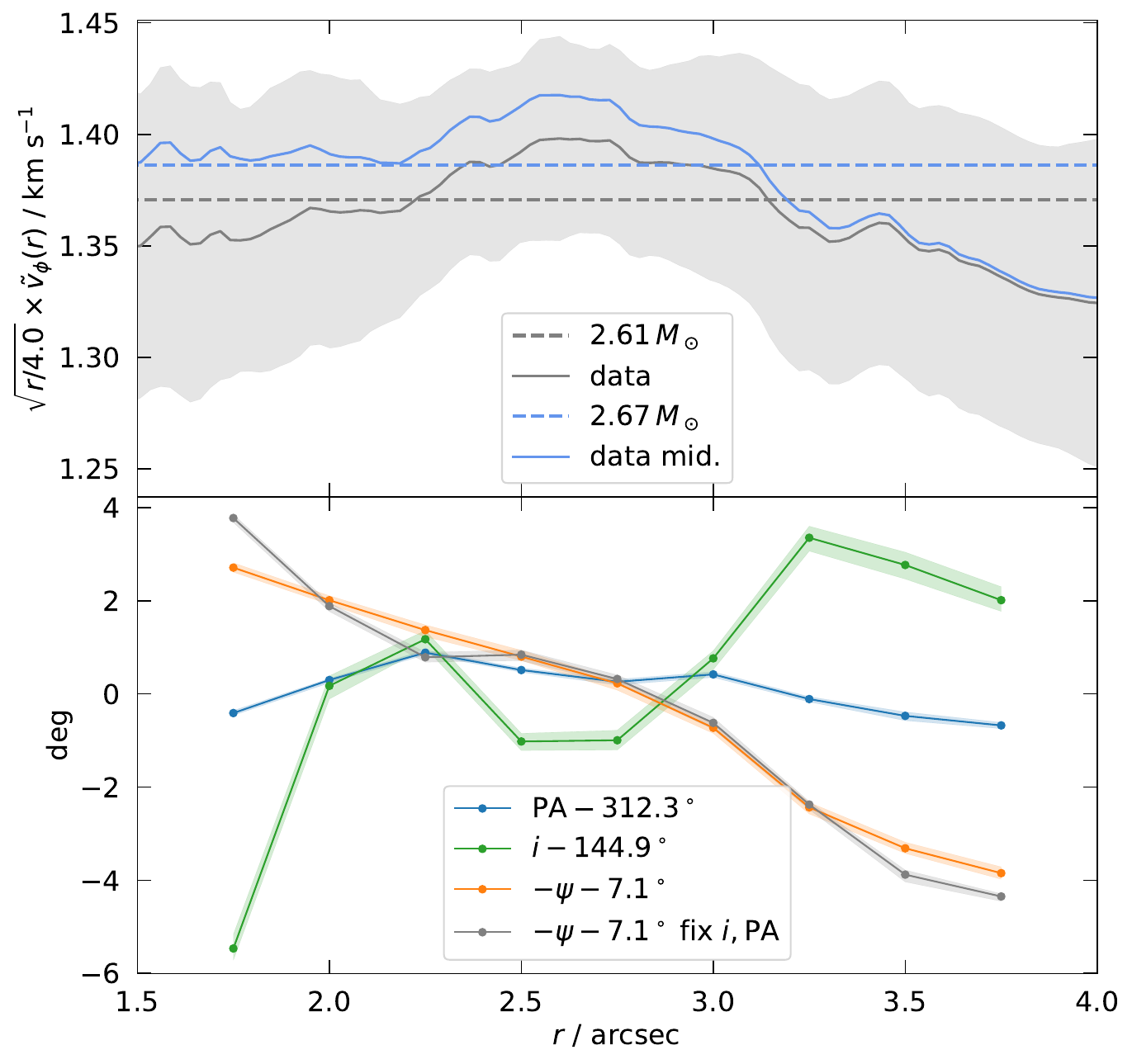}
\caption{Top: Non-axial velocity field $v_\circ - v_\circ^m$ in
  HD\,163296. Labels follow from Fig.\,\ref{fig:mom1_HD100546}d. The
  green circle is $0\farcs5$ in radius and is centered on the putative
  HD\,163296b. Bottom: Rotation curve and orientation profile, labels follow from
  Fig.\,\ref{fig:rotorient_HD100546}.
  \label{fig:HD163296}}
\end{figure}

Channel-map wiggles in $^{12}$CO have been identified in HD\,163296
\citep[][]{Pinte2018ApJ...860L..13P}, but we could not pick-up the
concomitant Doppler-flip in the velocity centroid in these and finer
angular resolution data \citep[based on the DSHARP large program data
  in ][see also their velocity
  centroid]{Isella2018ApJ...869L..49I}. In Fig.\,\ref{fig:HD163296},
the scatter within a radius of $0\farcs5$ centered at the location of
the putative protoplanet is $0.07$\,km\,s$^{-1}$, and a factor of
$\sim$4 smaller when smoothing to $\sim 50$\,au scales
\citep[corresponding to the section of the planetary wakes, see
  Fig.\,4 in][]{Pinte2018ApJ...860L..13P}. Thus the total amplitude of
any Doppler-flip at the predicted planet location is less than
$2\times0.05$\,km\,s$^{-1}$ (at 3$\sigma$).  The rotation curve reads
$\sim$1.8\,km\,s$^{-1}$ at the location of the putative protoplanet,
so a flip similar to that in HD\,100546 would have an amplitude of
$\sim$0.8\,km\,s$^{-1}$ and may escape detection.  We caution that the
spectral resolution of the DSHARP data is rather coarse,
$\sim$0.3\,km\,s$^{-1}$, so the CO line at the putative protoplanet
location is sampled with only 3 spectral points. This limits the
accuracy of the Gaussian centroid map.  We compute a systemic velocity
of $5.79\pm 0.02$\,km\,s$^{-1}$, and the global orientation is
$\psi_\circ= -7.1 \pm 2.2$\,deg, $i_\circ= 144.9 \pm 2.5$\,deg, and
PA$_\circ = 312.3 \pm 0.5$\,deg, but we note that this system may be
warped by $\sim 6$\,deg from $1\farcs5$ to $3\farcs5$.  The rotation
curve bears similar radial hydrostatic support modulations as
discussed by \citet[][]{Teague2018ApJ...860L..12T}.  For a distance of
101.5\,pc \citep[][]{Gaia2018A&A...616A...1G}, a Keplerian fit to
observed (surface) $\tilde{v}_\phi$ gives $M_\star = 2.61 \pm
0.08\,M_\odot$, and the midplane extrapolation gives $M_\star = 2.67
\pm 0.10\,M_\odot$, where the uncertainty are due to the hydrostatic
support modulations.








\begin{figure}
\centering
\includegraphics[width=\columnwidth]{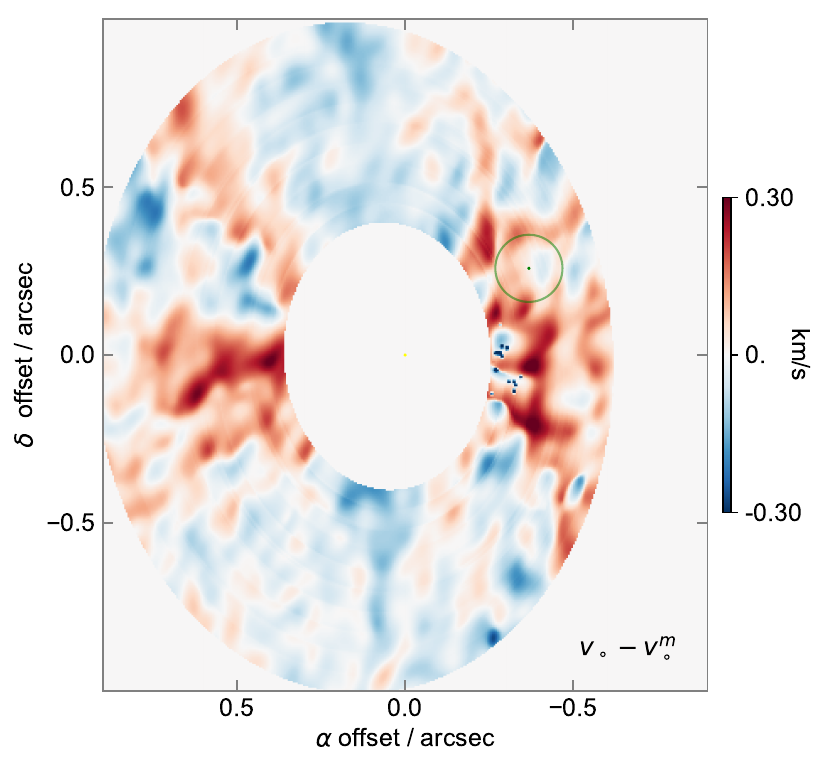}
\includegraphics[width=\columnwidth]{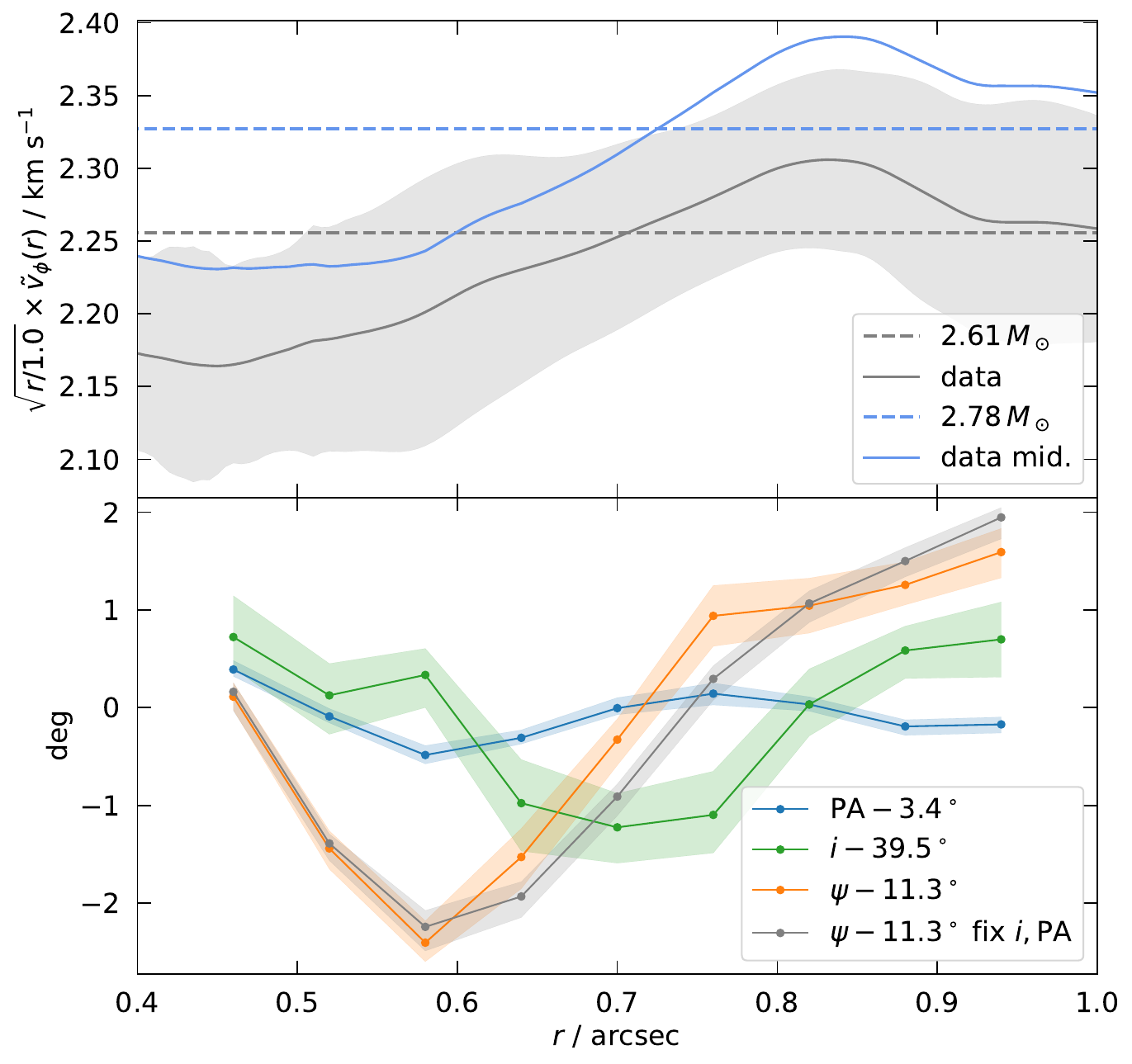}
\caption{Top: Non-Keplerian flow $v_\circ - v_\circ^m$ in HD\,97048.
  The green circle is centered on the putative HD\,97048b. Bottom:
  Rotation curve and orientation profile. Labels follow from
  Fig.\,\ref{fig:HD163296}.  \label{fig:HD97048}}
\end{figure}

We have also performed a similar analysis in HD\,97048, using the
$^{13}$CO(3-2) data from program {\tt 2016.1.00826.S}, \citep[][in
  this case the velocity sampling is
  0.1\,km\,s$^{-1}$]{Pinte2019arXiv190702538P}. The non-Keplerian flow
$v_\circ - v_\circ^m$ is shown in Fig.\,\ref{fig:HD97048}. Any
Doppler-flip in this ringed system is drowned by the systematics which
result in radial stripes, probably due to our neglect of the
far-side. The rotation curve reads $\sim$3\,km\,s$^{-1}$ at the
location of the putative protoplanet, so a Doppler-flip similar to
that in HD\,100546 would have an amplitude of
$\sim$1.3\,km\,s$^{-1}$. The systemic velocity is $v_s=
4.74\,$km\,s$^{-1}$, and the best fit orientation is $i_\circ= 39.5
\pm 0.75$\,deg, PA$=3.4 \pm 0.2$\,deg, $\psi_\circ= 11.3 \pm
1.3$\,deg. For a distance of 184.8\,pc
\citep[][]{Gaia2018A&A...616A...1G} the Keplerian mass from the midplane
extrapolation is $M_\star = 2.78 \pm 0.14\,M_\odot$.

%



\section{Conclusion} \label{sec:conclusions}

In an effort to indirectly detect embedded protoplanets using their
imprint on planet-disk interactions, we have designed a technique to
subtract the axially symmetric flow, or rotation curve, from velocity
centroid maps. Our goal was to pick-up protoplanets using the expected
Doppler-flip along the spiral wakes. Applications require
long-baseline and deep ALMA data. The case of HD\,163296, where a
protoplanet has previously been proposed based on disk kinematics,
yielded a negative result, limited by the measurement accuracy. Any
Doppler-flip in $^{12}$CO at the predicted planet location is less
than 0.05\,km\,s$^{-1}$ (at 3$\sigma$). Likewise in HD\,97048,
although these $^{13}$CO data may require a more refined averaging
procedure.  However, an application to the $^{12}$CO data in
HD\,100546 from paper\,I yielded a very conspicuous Doppler-flip,
whose properties match the expectations for a spiral wake in
planet-disk interactions in extension and in velocity sign, but not in
pitch-angle. Its coincidence with a fine ridge crossing a shallow
groove inside the continuum ring suggest a complex dynamical scenario,
and should inspire dedicated 3D gas+dust hydrodynamic simulations.



\acknowledgments

We thank the referee, Andrea Isella and Richard Teague for
constructive comments, as well as an inspiring discussion at a
workshop held in Monash University during July 2019.  Support was
provided by Millennium Nucleus RC130007 (Chilean Ministry of Economy),
FONDECYT grants 1171624 and 1191934, and by CONICYT-Gemini grant
32130007. This work used the Brelka cluster (FONDEQUIP project
EQM140101) hosted at DAS/U. de Chile.

\end{document}